# Transforming OPACs into Intelligent Discovery Systems: An AI-Powered, Knowledge Graph-Driven Smart OPAC for Digital Libraries

M. S. Rajeevan[#, *] and B. Mini Devi [#]
[#]Department of Library and Information Science, University of Kerala, Trivandrum, India.
Email: rajeevanms_2025@keralauniversity.ac.in

## Abstract

Traditional Online Public Access Catalogues (OPACs) are becoming less effective due to the rapid growth of scholarly literature. Conventional search methods, such as keyword indexing and Boolean queries, often fail to support efficient knowledge discovery. This paper proposes a Smart OPAC framework that transforms traditional OPACs into intelligent discovery systems using artificial intelligence and knowledge graph techniques. The framework enables semantic search, thematic filtering, and knowledge graph-based visualization to enhance user interaction and exploration. It integrates multiple open scholarly data sources and applies semantic embeddings to improve relevance and contextual understanding. The system supports exploratory search, semantic navigation, and refined result filtering based on user-defined themes. Quantitative evaluation demonstrates improvements in retrieval efficiency, relevance, and reduction of information overload. The proposed approach offers practical implications for modernizing digital library services and supports next-generation research workflows. Future work includes user-centric evaluation, personalization, and dynamic knowledge graph updates.

**Keywords**

## 1. Introduction

The rapid advancement of digital information has turned libraries into living hubs of knowledge, rather than traditional places to store printed materials. Historically, most online catalogues have used keyword indexing methods and conventional classification systems to allow for easy access to resources; however, these systems are becoming less able to meet users' expectations and have created negative experiences in finding information and satisfaction with the use of an OPAC. As digital collections grow larger and more complicated than ever, the need for intelligent systems capable of understanding a user's intent, building semantic relationships, and providing relevant results is now greater than ever before.

Machine Learning (ML) and Artificial Intelligence (AI) algorithms have been utilized to help overcome many of the constraints of Information Retrieval (IR) systems, including: expanded search capabilities; greater personalization of recommendations provided; and increased ability to automatically extract knowledge from large quantities of data/documents. Recent research indicates that by utilizing AI-powered semantic frameworks to develop IR capabilities within digital libraries, the precision of IR results and user engagement can be increased as a result of a better understanding of context and use of advanced representation models(Bi, 2025; Mala, 2024). KGs (knowledge graphs) are now used as a highly effective means of semantically describing relationships among different entities or things (e.g., documents, concepts, authors), and as such provide more robust navigation and discovery capabilities than classic indexing methods(Kroll et al., 2024; Mandal & Mandal, 2025). Knowledge graphs contain overall knowledge of how the content has developed in a particular area, and by this information create a model of complex interrelations between scholarly content that can enable systems to approximate human-like understanding of relationships between topics and relevance to queries.

The contextualized semantic retrieval systems that have been developed as an adaptive search technique incorporate user defined ontological and behavioral information along with graph-based representations which have been very effective solutions for improving both precision and recall within digital libraries(Bi, 2025; Shamsitdinova et al., 2024). The integration of artificial intelligence (AI) components such as embedding algorithms, keyword extraction, and interactive visualization tools extend a user's ability to intelligently navigate the digital library beyond simply searching result lists, thus enabling a greater degree of understanding of the thematic context relative to their queries(Ahmad, 2025; D'Souza, 2025).

Even with all these advancements made thus far, many libraries are still not utilizing AI and Knowledge Graph technology for their search systems. The purpose of this paper is to provide a comprehensive framework for converting OPACs into intelligent discovery systems based on applying AI-based semantic retrieval (e.g., semantic search) and visualizing knowledge graphs, while also using multiple sources of scholarly information (e.g., academic databases), as well as allowing users to filter based on the theme.

## 2. Review of Literature

Recent research has identified important progress and continuing challenges associated with the use of artificial intelligence (AI) and semantic technologies in digital libraries and library discovery systems. Traditional library discovery interfaces (OPACs) do not have the ability to interpret user intent or to model complex semantic relationships, significantly limiting their effectiveness as a means of finding information within large collections of diverse materials(Bi, 2025; Ignatowicz et al., 2025). Integrated Semantic Retrieval Frameworks (ISRFs) have been proposed to combine the construction of knowledge graphs, ontologies, and adaptive learning methods to provide better representation of both the content structure and the user's behaviour in digital libraries(Cruz et al., 2025; Niederhaus et al., 2025).

The primary representation of heterogeneous bibliographic items and their relations through knowledge graphs represents an area of growing interest in research and development; allowing more robust and rich semantic representation, enhanced interfaces for exploration beyond simple, textual lists of results, and more contextually based retrievals(Heidari et al., 2026; Rajeevan, Mini Devi, et al., 2025). A substantial amount of research and empirical evaluation within digital library systems supports the use of semantic graph structures to enable contextually based retrieval of information, personalized recommendations of resources for users, and to allow users to effectively navigate information that may contain multiple facets of information(Abu-Salih & Alotaibi, 2024; Rajeevan, B. Mini Devi, et al., 2025). Further research underscores how the utilization of knowledge graphs enriches the metadata of research papers, provides access to ontological reasoning tools, and allows for semantic searching that is more easily understood(D'Souza, 2025; Ling, 2025).

While advancements have been made, literature has identified a number of gaps regarding end-to-end frameworks which integrate multi-source retrieval, AI-based semantic processing, dynamic theme filtering, and interactive knowledge graph visualization in one OPAC. The majority of existing research has focused on either evaluating single components in isolation or evaluating only narrow-domain research; however, there is little empirical evidence regarding the effect of integrated systems on the reduction of information overload and the promotion of thematic discovery within all types of digital library settings. To fill these gaps, this study provides and assesses a Smart OPAC framework that integrates semantic retrieval, knowledge graph representation, and interactive filtering technologies to improve the discovery process of digital libraries.

## 3. Methodology

The purpose of this study is to design and develop a smart OPAC using an experimental research approach based on design science. The methodology used in developing this Smart OPAC combines the retrieval of scholarly data in real-time, the enrichment of content using semantics, and the visualization of materially integrated data from multiple sources into one cohesive system.

## 3.1 System Architecture and Workflow

The Smart OPAC system has 5 key sections that follow one after another to create the completed system: 1) User inputs query, 2) Multiple sources used to retrieve scholarly information, 3) Semantic Embedding and keyword extraction, 4) Filtering of results based on theme and filter criteria, and 5) Visualizing Knowledge Graphs as well as displaying interface for user. Streamlit is utilized to submit user inquiries on a web interface that enables parallel querying to three databases, Europe PMC, OpenAlex, and Semantic Scholar, using their respective APIs(Bi, 2025).

## 3.2 Semantic Representation and Keyword Extraction

To understand semantics in a way that surpasses simply matching without performing just a key word search, we utilize SBERT (Sentence-BERT) providing us with dense vector representations of paper title by the all-MiniLM-L6-2 embeddings(*SentenceTransformers Documentation*, 2026).

Given a document $d$, its embedding is computed as:

$$\mathbf{e}_d = SBERT(d)$$

KeyBERT uses representation of documents, in addition to a comparison of candidate phrases and the cosine similarity, to obtain a cosmetic keyword extraction(Fichman et al., 2025; Grootendorst, 2020/2026):

$$\text{score}(k, d) = \cos(\mathbf{e}_k, \mathbf{e}_d)$$

Only the top-ranked unigrams and bigrams are retained as themes, enabling interpretable semantic filtering.

## 3.3 Theme-Based Filtering Mechanism

The themes extracted from this collection will be presented in a dynamic manner through multi-select filters to allow for an interactive process of narrowing retrieved results. A paper will be retained if at least one of the extracted themes matches any of the semantic concepts that are currently selected(Elasticsearch, 2026; Kasenchak, 2019).

The relevance effectiveness of filtering is computed as:

$$Relevance(\%) = \frac{P_{\text{filtered}}}{P_{\text{retrieved}}} \times 100$$

where $P_{\text{retrieved}}$ denotes the total retrieved papers and $P_{\text{filtered}}$ represents papers retained after theme selection.

## 3.4 Knowledge Graph Construction and Visualization

A semantic representation of a query, resulting documents and themes is established through the creation of a knowledge graph (KG). The KG consists of vertexes for the user query, scientific paper(s) returned and theme(s) extracted from those papers; edges define the relationship between the query and scientific paper(s), and between scientific paper(s) and theme(s). The portrayal of the KG is achieved with the PyVis library(Bollegala et al., 2025; Gunel & Amasyali, 2023; Gupta et al., 2023; Zhang et al., 2024). This KG-driven representation enhances explainability and supports sense-making, addressing limitations of list-based OPAC interfaces(Chen et al., 2023).

## 3.4 Performance Evaluation Strategy

Performance of an information retrieval system is assessed based on the use of log-based metrics calculated from the execution data of the system, rather than from the relevance of the retrieved items. Average retrieval time per source is used to measure the efficiency of retrieval:

$$ART_s = \frac{1}{n}\sum_{i=1}^{n} t_i$$

Additionally, information overload reduction is quantified as:

$$Reduction(\%) = \left(1 - \frac{P_{\text{filtered}}}{P_{\text{retrieved}}}\right) \times 100$$

These metrics are appropriate for semantic OPACs that emphasize discovery and filtering over binary relevance assessment(Schopf et al., 2023).

## 3.5 Methodological Scope Note

Evaluation results will be presented independently. However, metrics definition, data logging, and analytical formulas are the important elements of the evaluation method and will be included here for purposes of replicability. Figure 1 presents the system workflow, illustrating system operation and data transformation.

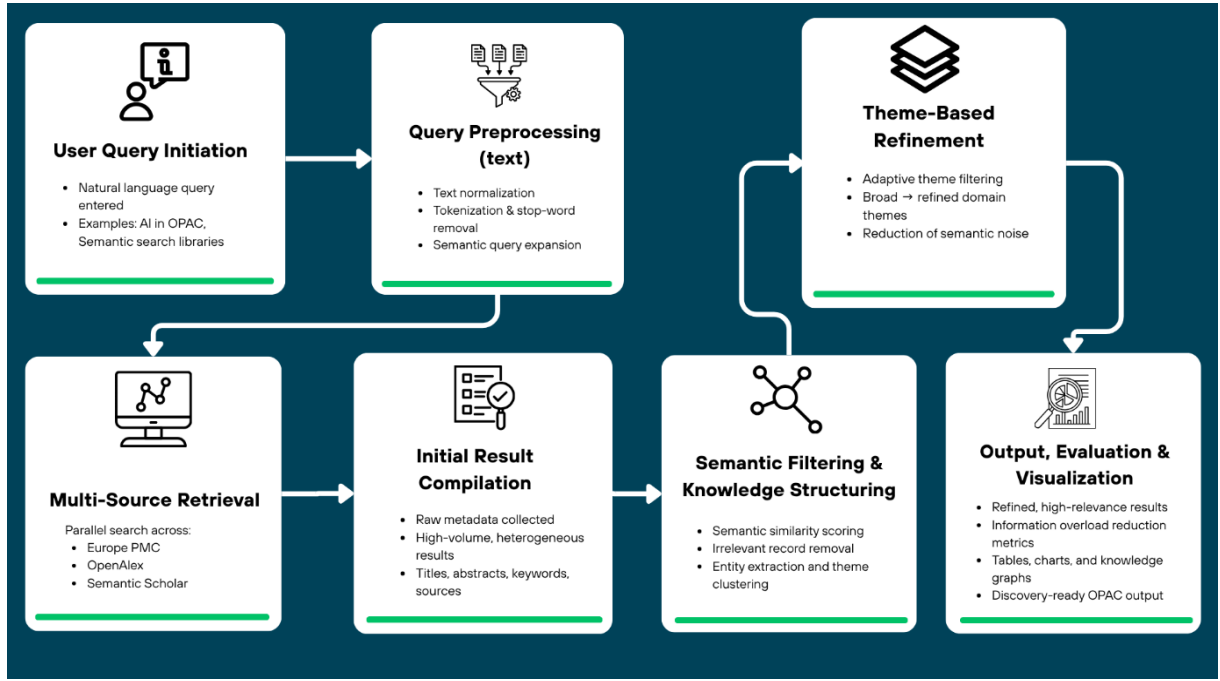

Figure 1: Workflow of the proposed semantic scholarly retrieval and filtering framework.

## 4. Evaluation and Result

In this section, we describe a thorough assessment of the proposed discovery framework based on semantic and knowledge graphs. Two types of evaluation methods have been used: quantitative methods (e.g., retrieval performance, relevance effectiveness, source contribution, and reduction of information overload) and qualitative methods (e.g., knowledge graph visualizations to demonstrate semantic structure and the relatedness of themes among retrieved literature).

### 4.1 Retrieval Performance Across Scholarly Sources

The data displayed in Table 1 includes summary statistics of the average time taken to retrieve an item, total number of items retrieved and number of queries processed for each of the three main data sources: Europe PMC, OpenAlex and Semantic Scholar.

Table 1: Retrieval Performance by Source.

| Sl. No. | Source | Avg Retrieval Time(sec) | Total Papers Retrieved | Queries Handled |
|---|---|---|---|---|
| 1 | Europe PMC | 1.318 | 50 | 5 |
| 2 | OpenAlex | 3.646 | 50 | 5 |
| 3 | Semantic Scholar | 1.124 | 10 | 1 |

The results show that the fastest average retrieval times were achieved by Semantic Scholar (1.124 sec) and Europe PMC (1.318 sec) making them good candidates for time-critical scholarly discovery activities. Conversely, OpenAlex has a much longer average retrieval time (3.646 sec) indicating a greater amount of latency.

Although Europe PMC and OpenAlex retrieved more overall paper data than Semantic Scholar (50 per query set), there were notable differences in query response times (latency). The higher total volume of papers retrieved (Europe PMC and OpenAlex) demonstrates their infrastructure enables more comprehensive index coverage and repository capabilities. Semantic Scholar did produce results for only one query. The lower total number of papers retrieved (10) is indicative of a more selective and circumstance-driven download method than that based on the original framework's download protocol. Thus, this is not an indication of a limitation of the original framework.

The visual depiction of retrieval speed in Figure 2 provides another illustration of how average retrieval times vary across three repository sources and demonstrates the significant difference in retrieval latencies between OpenAlex and the two source repositories, multi-source aggregation amasses a balance between speedy retrieval and comprehensive scholarly access due to trade-offs between these two characteristics.

Overall using heterogeneous databases in a Smart OPAC framework is a successful design decision.

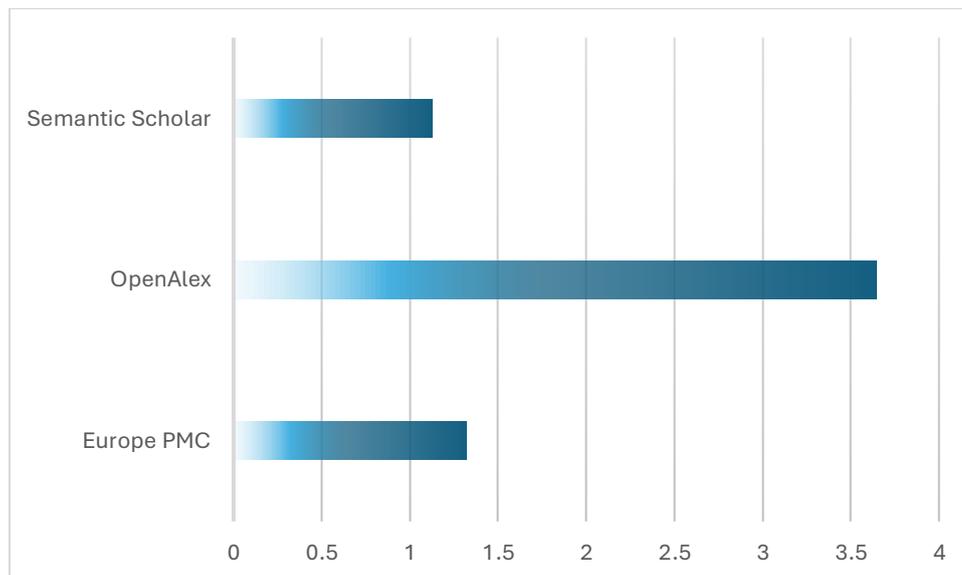

Figure 2: Average Retrieval Time Across Scholarly Data Sources.

## 4.2 Query-Level Relevance Effectiveness

The results of Table 2 show significant variation in how effective the process of filtering out non-relevant items has been for each query type according to how specific that query is semantically.

Table 2: Relevance Effectiveness by Query.

| Sl. No. | Query | Total Papers Retrieved | Papers After Filter | Relevance % |
|---|---|---|---|---|
| 1 | AI in OPAC | 20 | 6 | 30 |
| 2 | Information retrieval LIS | 20 | 8 | 40 |
| 3 | knowledge graph in digital libraries | 20 | 13 | 65 |
| 4 | library recommendation systems | 20 | 13 | 65 |
| 5 | semantic search libraries | 30 | 25 | 83.33 |

Query types that are well-defined and good at capturing lots of meaning seemed to be slightly higher than the others. Examples include: "semantic search libraries" at 83.33% and "knowledge graph in digital libraries" at 65%. However, when looking at a broad query such as "AI in OPAC," only 30% retained relevance. This suggests that queries that are broad suffer from ambiguity and therefore did not produce retainable relevance precision.

These results indicate that the use of semantic filtering, coupled with a knowledge graph, is highly beneficial for those under well-defined concepts, in which domain entities and their relationships are explicitly defined. As more ambiguous queries are completed, semantic noise will be introduced into the returned document set, resulting in a decrease in the amount of relevant document content available for retention. The proposed framework increases the precision of the system by eliminating irrelevant publication records, while providing a large amount of scholarly material of high quality, especially in domains where the conceptual boundaries are clear-cut.

**4.3 Source Contribution Analysis**

Table *3* provides the proportion of each of the retrieval products made by each of the data sources, where Europe PMC and OpenAlex each constituted 45.45% of the total retrieved collection, while the remaining 9.09% came from Semantic Scholar.

Table 3: Source Contribution Distribution.

| Sl. No. | Source | Papers Contributed | Contribution % |
|---|---|---|---|
| 1 | Europe PMC | 50 | 45.45 |
| 2 | OpenAlex | 50 | 45.45 |
| 3 | Semantic Scholar | 10 | 9.09 |

The balanced distribution of contributions represents a complementarity between biomedical and multidisciplinary resources. Biomedical-oriented repositories such as Europe PMC complement the multidisciplinary source such as Open Alex. In contrast, Semantic Scholar contains only selectively produced but semantically rich content. The distribution supports the project's design decision to include heterogeneous scholarly sources in order to enhance both overall coverage and diversity of the domains covered.

## 4.4 Information Overload Reduction

Table *4* shows the effectiveness of the Smart OPAC model, which was analysed by looking at how many papers were found before applying semantic filtration as compared to how many remained following relevance-oriented filtration and theme-conducted filtration. The percentage drop indicates the degree to which marginally relevant or redundant documents were removed thus benefiting users' attention and cognitive efficiency.

Table 4: Information Overload Reduction by Query and Source.

| Sl. No. | Query | Source | Papers Retrieved | Papers After Filter | Reduction % |
|---|---|---|---|---|---|
| 1 | library recommendation systems | Europe PMC | 10 | 3 | 70 |
| 2 | library recommendation systems | OpenAlex | 10 | 10 | 0 |
| 4 | knowledge graph in digital libraries | Europe PMC | 10 | 4 | 60 |
| 5 | knowledge graph in digital libraries | OpenAlex | 10 | 9 | 10 |
| 7 | AI in OPAC | Europe PMC | 10 | 1 | 90 |
| 8 | AI in OPAC | OpenAlex | 10 | 5 | 50 |
| 10 | semantic search libraries | Europe PMC | 10 | 6 | 40 |
| 11 | semantic search libraries | OpenAlex | 10 | 9 | 10 |
| 12 | semantic search libraries | Semantic Scholar | 10 | 10 | 0 |
| 13 | information retrieval LIS | Europe PMC | 10 | 1 | 90 |
| 14 | information retrieval LIS | OpenAlex | 10 | 7 | 30 |

The research revealed a marked decline in the number of overloads for complicated concepts and queries that cross traditional institutional boundaries (e.g., "AI in OPAC" and "Information Retrieval LIS") when evaluated using Europe PMC. Both of these examples show an approximate 90 percent reduction; therefore, they demonstrate how effectively the semantic filtering and knowledge graph theme selection methods can identify relevant information from a larger set of search results. Europe PMC's superior performance is due in large part to its integration of a variety of biomedical/life science repositories into one database. This results in Europe PMC containing richer metadata, structured abstracts, and standardised subject annotations than any other source. These features have a positive impact on entity extraction and semantic similarity measurement, thereby enabling better discrimination of relevancy when performing post-retrieval filtering. Additionally, there were significant reductions (> 70 percent and > 60 percent, respectively) for "Library Recommendation Systems" and "Knowledge Graph in Digital Libraries."

In comparison, the OpenAlex dataset has a lower rate of reduction for some queries such as "library recommendation systems" (0%) and "knowledge graph in digital libraries" (10%). This indicates that repository-level characteristics, specifically the structure of the metadata and the level of abstraction, significantly affect the efficiency of semantic filtering after retrieval.

Semantic Scholar exhibits 0% retrieval reduction for the query "semantic search libraries" of retrieved articles, indicating no documents were excluded from the results after filtering occurred. This illustrates that although initial precision is very good; the selection of documents for inclusion is not representative of the actual population, but rather indicative of a limitation in the document retrieval process itself. The selective nature of Semantic Scholar's involvement is attributed to its query-dependent document retrieval method, and the function of its API to limit available result sets to specific topic definitions.

The overall results indicate that Smart OPAC results eliminate information overload while still retaining topical relevance, especially for requests with high semantic ambiguity. As shown in Figure 3, the average reduction of different LIS query types were illustrated graphically to show the overall average percent reductions to these types of queries. Semantically more complicated topics, such as "AI in OPAC" and "Information Retrieval LIS," have larger average percent reductions than semantically simpler topics with less precision, such as "Semantic Search Libraries."

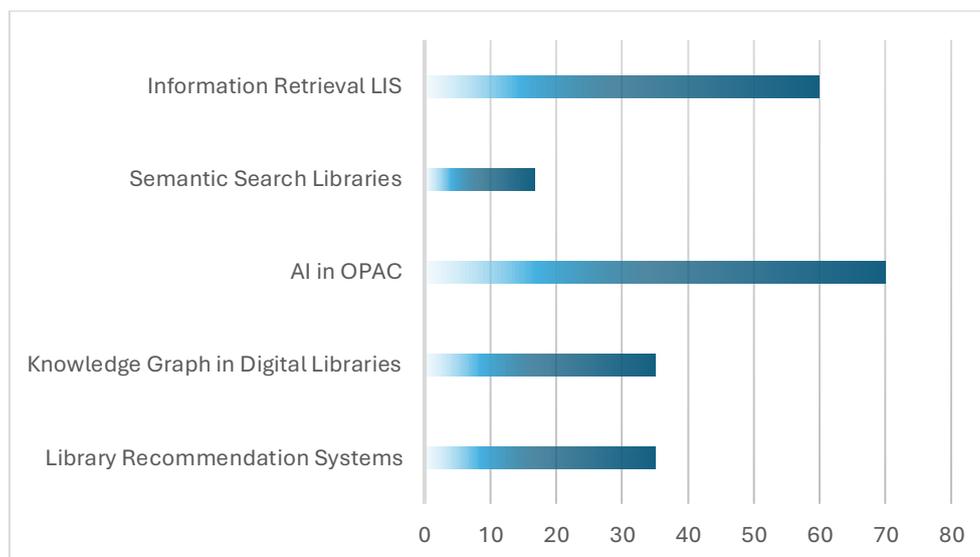

Figure 3: Average information overload reduction by query type.

### 4.5 Knowledge Graph–Based Discovery Insights

In addition to the quantitative evaluation, three Knowledge Graphs (visualizations) were produced, to analyse semantic structure, thematic clumping, and interdisciplinary relationships in the extracted corpus.

The Figure 4 depicts an initial semantic environment that reflects the semantic landscape of the unfiltered data before thematic selection has occurred. The knowledge graph in digital libraries is shown to relate to many different domains and has many interconnected themes, including; adaptive semantic retrieval, digital humanities, data management, ontology-based knowledge, oncology intelligence, and benchmarking knowledge. Thus, this visual representation of the large quantity of interdisciplinary research signals captured by the system over an extended time period demonstrates the availability of interdisciplinary research signals on a large scale.

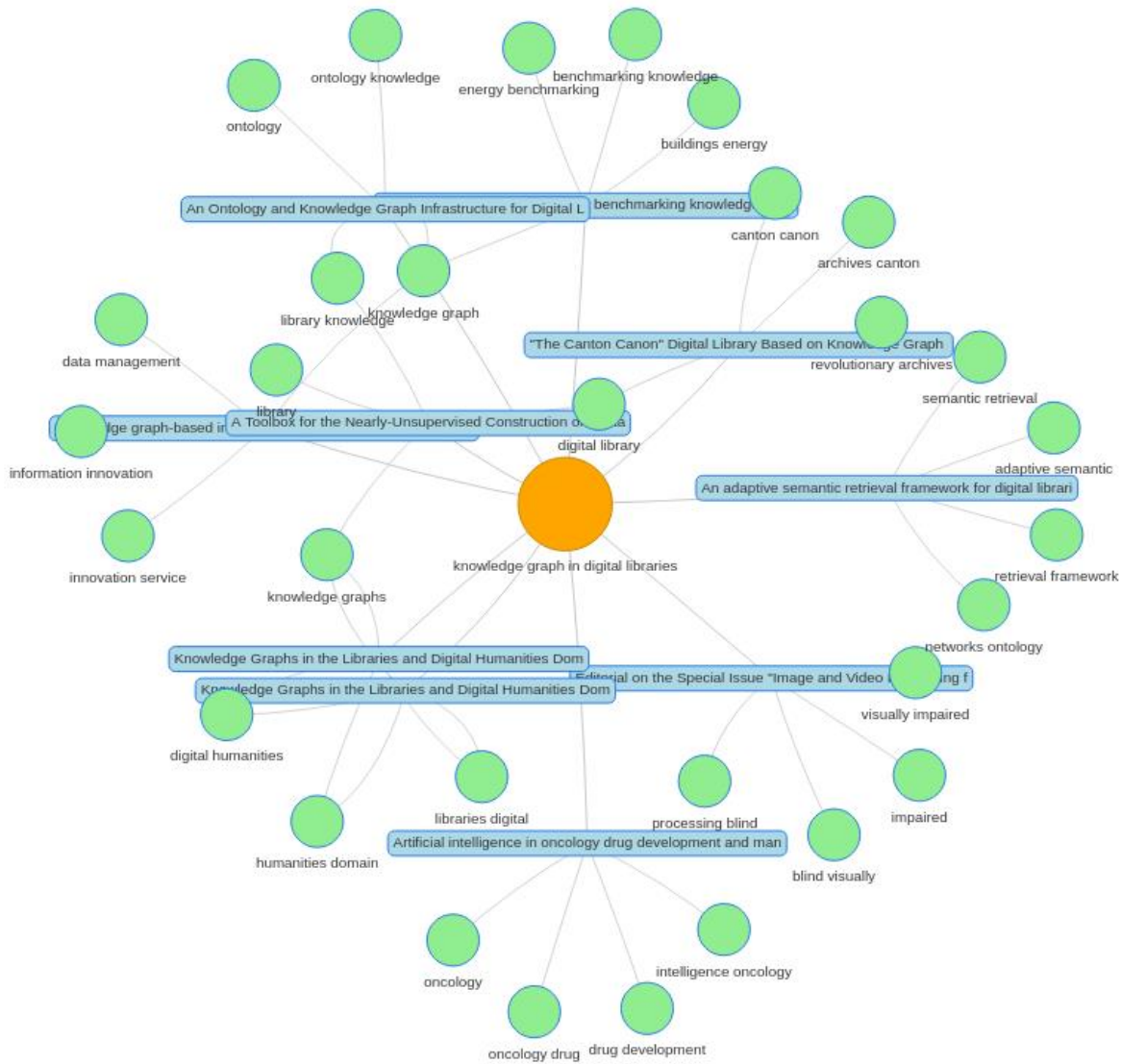

Figure 4: Knowledge Graph 1 (Unfiltered Theme Space)

The second iteration depicted in Figure 5 concentrates on developing an understanding of the following key themes or topic areas: adaptive semantic benchmarking knowledge, building energy use, data management, and digital humanities. The overall structure of the graph is clearer and has less network "noise", demonstrating how filtering based on themes improves clarity of thought and conceptual boundaries while keeping relevant cross-domain connections.

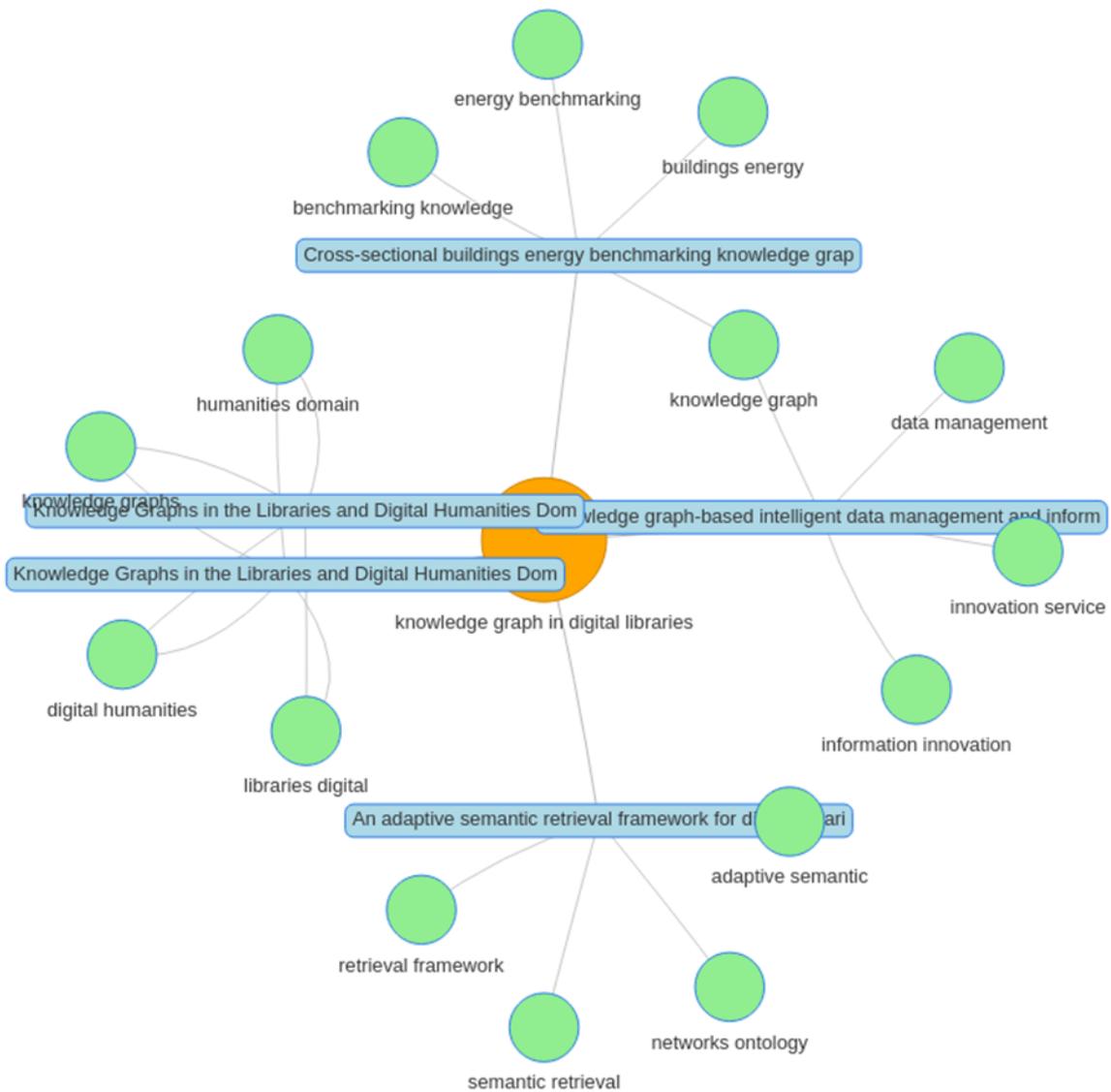

Figure 5: Knowledge Graph 2 (Filtered Semantic Themes)

The Figure 6 also has an isolation of more advanced thematic clusters, i.e., intelligence oncology, information innovation, networks ontology, retrieval frameworks and semantic retrieval. The existence of a large number of tightly interconnected subgraphs serves as evidence that the knowledge graph can effectively reveal a vast array of deep semantic relationships and facilitate advanced exploratory searching and knowledge discovery.

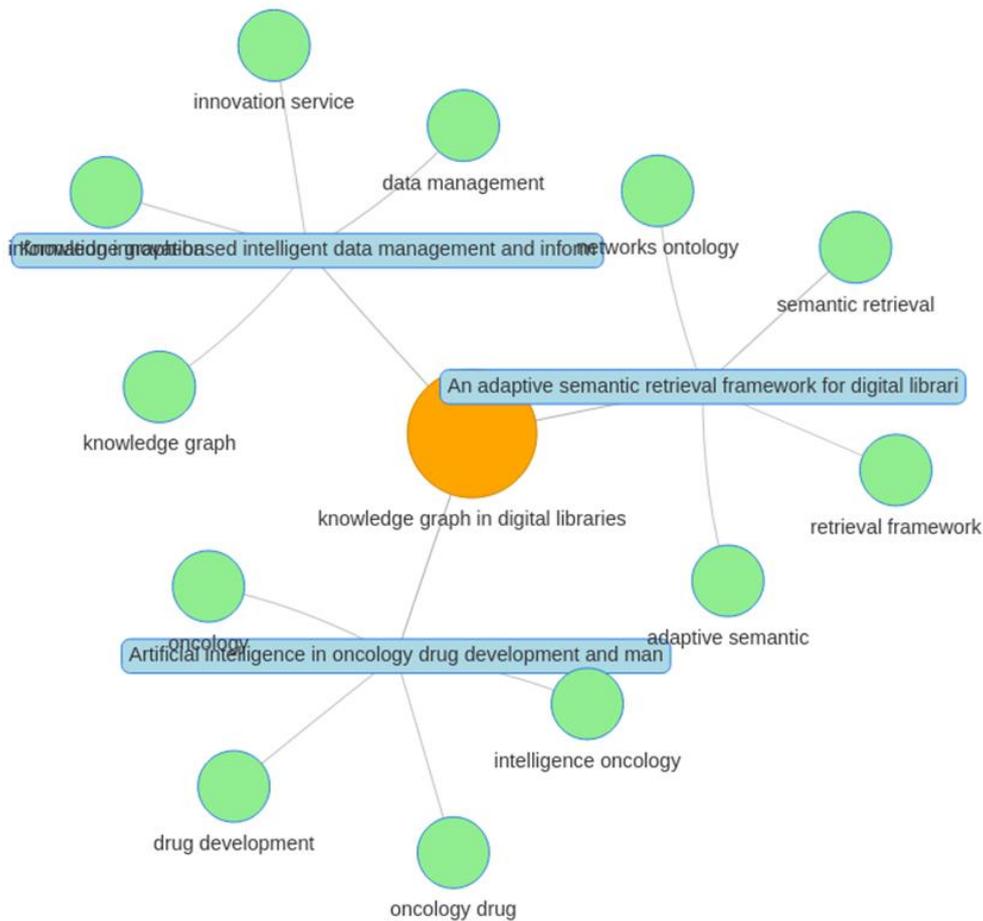

Figure 6: Knowledge Graph 3 (Filtered Semantic Themes)

These visualizations also support the conclusion that knowledge graph retrieval improves the relevance of results and decreases the volume of irrelevant results and, further, provides the ability to create knowledge at a higher level of meaning than was possible with OPAC keyword-based systems alone.

### 4.5 Summary of Evaluation Findings

In general, the assessment indicates that the proposed system:

- Enhances information discovery by providing multi-source orchestration to improve search retrieval performance
- Provides improved relevance through filtering based on semantics and knowledge graphs
- Reduces significantly the problem of information overload
- Supports rich semantic exploration and interdisciplinary knowledge discovery

These results validate the proposed system as an effective next-generation resource discovery framework for intelligent OPACs and digital libraries.

## 5. Discussion

Results of the evaluation indicate that the Smart OPAC framework successfully mitigates the traditional library discovery system's limitations by enhancing retrieval efficiency, increasing thematic relevancy of results, and enabling knowledge-driven exploration. The comparative analysis found that Europe PMC and Semantic Scholar provide faster average retrieval times than OpenAlex, indicating that the selection of source impacts the responsiveness of the system.

By conducting a query level relevance analysis, it was determined that theme based filtering significantly improves the quality of the results produced. After utilizing the filtering technique, the queries of "semantic search libraries" and "knowledge graph in digital libraries" received high levels of relevance indicating that semantic embeddings and theme based constraints are very beneficial for conceptually rich topics (LIS). The findings around the reduction of information overload further validate the importance of the proposed model. The overall high amount of reduction, particularly for queries such as: AI in OPAC and Information Retrieval LIS demonstrate that the model was able to successfully remove low marginally relevant or marginally related documents while keeping all core literature.

Three different visualizations of knowledge graph demonstrate that the Smart OPAC can now provide better qualitative help to users in their discovery efforts. The first graph is created without any theme selection; therefore, it represents all the diverse and disorganized conceptual content that was retrieved by the system. The second and third graphs represent more organized and clustered sets of concepts as a result of selective theme choices made by the user, which enables them to explore the subject matter more thoroughly in specific research areas.

In conclusion, the results imply that by combining semantic retrieval, theme-based filtering and knowledge graph visualisation there is potential for significant increases in OPAC performance; however, beyond improving performance metrics, the new system facilitates more profound and direct intellectual engagement with the literature in ways that correspond to the evolving expectations of next generation digital library services.

## 6. Conclusion and Future Work

This research project illustrated how conventional Online Public Access Catalogs (OPACs) can change form into smart technological tools for discovering knowledge through use of artificial intelligence-powered retrieval algorithms, thematic content filtering and knowledge graph-based opportunities for finding new scholarly information. The overall results demonstrated the Smart OPAC model substantially increases relevance of retrieved materials, decreases the amount of irrelevant material returned, and enables users to gain a greater understanding of the semantic content across multiple formats and types. The combination of both illustrates that the model is capable of transcending keyword-based search to provide context-sensitive or concept-based access to information from digital libraries.

While this solution is effective; its current implementation has some limitations. The evaluation was performed with a limited number of queries and data sources and user-centered usability tests were not included. Future work will include expanding the coverage of the data sources used, adding analytics for user interaction, and implementing personalization based on user input and feedback. Further enhancements involve ongoing addition of semantic values created via dynamic ontology enrichment, real-time updating of the knowledge graph, and advanced analytical techniques applied to develop recommendations and trends using graph-based analysis. Other enhancements represent additional capabilities available to Smart Online Public Access Catalogs that will help make them smart systems for knowledge discovery that can support advanced research and support decision-making in the digital library environment.

**Author Biographies**

**M. S. Rajeevan** is a Research Scholar from the Department of Library and Information Science in University of Kerala, India. His research focuses on semantic retrieval and its applications in knowledge discovery and intelligent systems. Rajeevan has conceptualised the Smart OPAC framework, implemented the multi-source retrieval pipeline, carried out data collection, tested the data, created tables and visualisations and written the manuscript for this study.
**ORCID iD**: https://orcid.org/0009-0009-3320-3078
**Website**: https://msrajeevan.in/

**Dr. B. Mini Devi** is currently an Assistant Professor and Director of the Centre for Information Literacy Studies at the University of Kerala, where she previously served as Head of the Department. She earned her PhD from the Cochin University of Science and Technology and has substantial experience in LIS Research and Academic Leadership. In addition, Mini Devi has provided conceptual guidance; supervised the research design and evaluation framework; refined study objectives; and critically reviewed the manuscript.
**IRINS**: https://keralauniversity.irins.org/profile/175002
**Website**: https://bminidevi.com/

**References**

1. Abu-Salih, B., & Alotaibi, S. (2024). A systematic literature review of knowledge graph construction and application in education. *Heliyon*, *10*(3), e25383. https://doi.org/10.1016/j.heliyon.2024.e25383

2. Ahmad, Y. A. A. (2025). Examining The Role Of Artificial Intelligence And Machine Learning In Improving Search Functionality And Metadata Organization In Digital Libraries. *International Journal On Digital Libraries (IJDL)*, *2*(1), 1–7. https://iaeme.com/Home/article_id/IJDL_02_01_001

3. Bi, R. (2025). An adaptive semantic retrieval framework for digital libraries integrating graph neural networks, ontology, and user behavior. *Scientific Reports*, *15*(1), 40528. https://doi.org/10.1038/s41598-025-24276-1

4. Bollegala, D., Otake, S., Machide, T., & Kawarabayashi, K.-I. (2025). A Metric Differential Privacy Mechanism for Sentence Embeddings. *ACM Trans. Priv. Secur.*, *28*(2), 20:1-20:34. https://doi.org/10.1145/3708321

5. Chen, B., Chen, K., Yang, Y., Amini, A., Saxena, B., Chávez-García, C., Babaei, M., Feizpour, A., & Varró, D. (2023). *Towards Improving the Explainability of Text-based Retrieval with Knowledge Graphs* (arXiv:2301.06974). arXiv. https://doi.org/10.48550/arXiv.2301.06974

6. Cruz, T. da, Tavares, B., & Belo, F. (2025). *Ontology Learning and Knowledge Graph Construction: A Comparison of Approaches and Their Impact on RAG Performance* (arXiv:2511.05991). arXiv. https://doi.org/10.48550/arXiv.2511.05991

7. D'Souza, J. (2025). Taming the Generative AI Wild West: Integrating Knowledge Graphs in Digital Library Systems. *The Code4Lib Journal*, (60). https://journal.code4lib.org/articles/18277

8. Elasticsearch. (2026). *What is Semantic Search? | A Comprehensive Semantic Search Guide*. Elastic.Co. https://www.elastic.co/what-is/semantic-search

9. Fichman, N., Isaacson, H., & Vanetik, N. (2025). A New Dataset for Keyword Extraction from IT Job Descriptions. *Advances in Information Retrieval: 47th European Conference on Information Retrieval, ECIR 2025, Lucca, Italy, April 6–10, 2025, Proceedings, Part III*, 381–390. https://doi.org/10.1007/978-3-031-88714-7_37

10. Grootendorst, M. (2026). *MaartenGr/KeyBERT* [Python]. https://github.com/MaartenGr/KeyBERT (Original work published 2020)

11. Gunel, K., & Amasyali, M. F. (2023). Boosting Lightweight Sentence Embeddings with Knowledge Transfer from Advanced Models: A Model-Agnostic Approach. *Applied Sciences*, *13*(23), 12586. https://doi.org/10.3390/app132312586

12. Gupta, V., Dixit, A., & Sethi, S. (2023). An Improved Sentence Embeddings based Information Retrieval Technique using Query Reformulation. *2023 International Conference on Advancement in Computation & Computer Technologies (InCACCT)*, 299–304. https://doi.org/10.1109/InCACCT57535.2023.10141788

13. Heidari, G., Stocker, M., & Auer, S. (2026). Enhancing Information Retrieval in Digital Libraries Through Unit Harmonisation in Scholarly Knowledge Graphs. In S. Oh, A. Doucet, M. Buranarach, I. Buenrostro-Cabbab, Y. Liu, & B. S. Olgado (Eds.), *Intelligence and Equity: Shaping the Future of Knowledge* (pp. 170–184). Springer Nature. https://doi.org/10.1007/978-981-95-4861-3_15

14. Ignatowicz, J., Kutt, K., & Nalepa, G. J. (2025). *Position Paper: Metadata Enrichment Model: Integrating Neural Networks and Semantic Knowledge Graphs for Cultural Heritage Applications* (arXiv:2505.23543). arXiv. https://doi.org/10.48550/arXiv.2505.23543

15. Kasenchak, R. T. (2019). What is Semantic Search? And why is it important? *Information Services and Use*, *39*(3), 205–213. https://doi.org/10.3233/ISU-190045

16. Kroll, H., Pirklbauer, J., Kalo, J.-C., Kunz, M., Ruthmann, J., & Balke, W.-T. (2024). A discovery system for narrative query graphs: Entity-interaction-aware document retrieval. *International Journal on Digital Libraries*, *25*(1), 3–24. https://doi.org/10.1007/s00799-023-00356-3

17. Ling, C. (2025). Integrating knowledge graphs into academic libraries: A comparative study of Sub-Saharan and Chinese education systems. *Information Development*, 02666669251389139. https://doi.org/10.1177/02666669251389139

18. Mala, J. M. (2024). From Dewey to Deep Learning: Exploring the Intellectual Renaissance of Libraries through Artificial Intelligence. *Journal of Information and Knowledge*, 29–38. https://doi.org/10.17821/srels/2024/v61i1/171001

19. Mandal, P. S., & Mandal, S. (2025). *Empowering Digital Library Management through Knowledge Graphs*. 379–393. https://doi.org/10.2991/978-2-38476-533-1_23

20. Niederhaus, M., Migenda, N., Weller, J., Kohlhase, M., & Schenck, W. (2025). Integrating Graph Retrieval-Augmented Generation into Prescriptive Recommender Systems. *Big Data and Cognitive Computing*, *9*(10), 261. https://doi.org/10.3390/bdcc9100261

21. Rajeevan, M. S., Mini Devi, B., Anoop, V. S., & Mallikarjuna, C. (2025). Mapping of Emerging Technological Trends in Library and Information Science: A Computational Approach Using Sentiment Analysis, Topic Modeling and Network. *DESIDOC Journal of Library & Information Technology*, *45*(4), 326–338. http://publicationsdrdo.in/index.php/djlit/article/view/21029

22. Rajeevan, M. S., Mini Devi, Mallikarjuna, C., & V. S., A. (2025). *NLP-Driven Knowledge Extraction and Thematic Classification of Translated Ancient Indian Medical Texts* (SSRN Scholarly Paper 5217974). Social Science Research Network. https://doi.org/10.2139/ssrn.5217974

23. Schopf, T., Gerber, E., Ostendorff, M., & Matthes, F. (2023). AspectCSE: Sentence Embeddings for Aspect-based Semantic Textual Similarity Using Contrastive Learning and Structured Knowledge. *Proceedings of the Conference Recent Advances in Natural Language Processing - Large Language Models for Natural Language Processings*, 1054–1065. https://doi.org/10.26615/978-954-452-092-2_113

24. *SentenceTransformers Documentation*. (2026). https://sbert.net/

25. Shamsitdinova, M., Khashimova, D., Niyazova, N., Nasirova, U., & Khikmatov, N. (2024). Harnessing AI for Enhanced Searching in Digital Libraries: Transforming Research Practices. *Indian Journal of Information Sources and Services*, *14*(3), 102–109. https://doi.org/10.51983/ijiss-2024.14.3.14

26. Zhang, B., Chang, K., & Li, C. (2024). *Simple Techniques for Enhancing Sentence Embeddings in Generative Language Models* (arXiv:2404.03921). arXiv. https://doi.org/10.48550/arXiv.2404.03921